\documentclass[prd,preprint,double-spaced,superscriptaddress,showpacs]{revtex4-1}
\begin{document}

\title{Finding a Spherically Symmetric Cosmology from Observations in Observational 
Coordinates -- Advantages and Challenges}

\date{\today}

\author{M.E. Ara\'{u}jo}
\affiliation{ Departamento de F\'{\i}sica-Matem\'{a}tica, Instituto de F\'{\i}sica,\\
    Universidade Federal do Rio de Janeiro,\\
            21.945-970, Rio de Janeiro, R.J., Brazil}
\author{W.R. Stoeger}
\affiliation{ Vatican Observatory Research Group \\
      Steward Observatory, University of Arizona, \\
      Tucson, AZ 85721, USA}

\begin{abstract}

One of the continuing challenges in cosmology has been to determine the large-scale
space-time metric from observations with a minimum of assumptions -- without,
for instance, assuming that the universe is almost Friedmann-Lema\^{i}tre-Robertson-Walker
(FLRW). If we are lucky enough this would be a way of demonstrating that our universe is
FLRW, instead of presupposing it or simply showing that the observations are consistent
with FLRW. Showing how to do this within the more general spherically symmetric, inhomogeneous space-time framework takes us a long way towards fulfilling this goal. In recent work researchers have shown how this can be done both in the traditional Lema\^{i}tre-Tolman-Bondi (LTB) 3 + 1 coordinate framework, and in the observational coordinate (OC) framework, in which the radial coordinate $y$ is null (light-like) and measured down the past light cone of the observer.

In this paper we investigate the stability of solutions, and the use of data in the OC field 
equations including their time evolution -- i.e. our procedure is not restricted to our past light 
cone -- and compare both approaches with respect to the singularity problem at the maximum of the angular-diameter distance, the stability of solutions, and the use of data in the field equations. We also compare the two approaches with regard to determining the cosmological constant $\Lambda$. This allows a more detailed account and assessment of the OC integration procedure,
and enables a comparison of the relative advantages of the two equivalent solution
frameworks. Both formulations and integration procedures should, in principle, lead to 
the same results. However, as we show in this paper,  the OC procedure manifests certain advantages, particularly in the avoidance of coordinate singularities at the maximum of the
angular-diameter distance, and in the stability of the solutions obtained. This particular feature 
is what allows us to do the best fitting of the data to smooth data functions and the possibility of constructing analytic solutions to the field equations. Smoothed data functions enable us to 
include properties that data must have within the model. \\

\end{abstract}


\maketitle

\section{Introduction}

Over the last 15 years, there has been great progress made in cosmology. The concordance
model -- a flat FLRW universe with $\Omega_m \approx 0.27$ and $\Omega_{\Lambda} \approx
0.73$, where these $\Omega$'s represent the matter content and dark-energy content of the
universe, respectively -- has received continuing strong support from an explosion of detailed observational
data together with its careful analysis vis-a-vis theoretical considerations. This model seems
to provide an increasingly reliable account of all the observations relevant to the structure
and history of our universe. \\

However, there are lingering uncertainties and unresolved controversies -- particularly with
regard to dark-energy (is it simply the cosmological constant $\Lambda$, something dynamical, evolving, or an alias for the influence of intermediate and large-scale inhomogeneities?) and with regard to the overall adequacy of perturbed FLRW models in a very inhomogeneous universe, at least on small and intermediate length scales. It would be reassuring to actually demonstrate from observations that the metric of the universe on scales larger than a certain definite value -- or averaged over volumes larger than a certain definite value -- is almost FLRW, instead of simply assuming that, and then finding that the observations are more or less consistent with almost-FLRW. One of the difficulties, of course, is that they might also be consistent with one or more cosmologies which are far from FLRW. At present we have no assurance that an almost-FLRW universe is the unique best-fit model to the observations, even on the largest scales. We do not even know reliably what is the smallest scale on which the universe is almost-FLRW. And it is obvious that it certainly is not on small and intermediate scales.\\

There are certainly strong indications that our universe is almost-FLRW, as already emphasized. But in order for these  to constitute definitive evidence more is needed -- in particular either observational evidence  that the cosmological principle (Copernican principle) is correct, that we are not located in a large scale under-dense bubble, for instance, or that observational data constraining the models is almost-FLRW. For, a careful critical treatment of these issues, see \cite{CM}. As they emphasize, any such conclusions demand non-perturbative investigations and confrontation with the data. Though some recent studies \cite{CS,CFL,ZS} argue that data on spectral distortions due to scattering of CMB photons off free electrons between  $z=1$ and $z=10$ significantly restrict the kind of void which may surround us, more precise and extensive measurements together with non-perturbative modeling beyond FLRW is essential, as Clarkson and Maartens \cite{CM} stress.\\

In order to begin to provide a broader framework for eventually banishing these uncertainties -- and 
possibly providing confirmation of the large-scale almost-FLRW character of our universe -- there
has been considerable recent theoretical work oriented towards showing how we can begin from
an assumed cosmological ansatz considerably more general than FLRW, and determine its metric from observations in a rigorous way. This is possible in principle \cite{Ellis et al}, though
meshing the observational and the theoretical has proven very tricky, and accumulating enough
precision data of the right sorts is still uncertain. In particular, a number of workers,
including ourselves, \cite{OC III, AS, ASR, AABFS, ASAB, AS2009b, AS2010a, mustaph:hell, liu:hell, mcclur:hell, hell:alna, MaaMat, InhomUniv, Hellamulti} have begun with a general spherically symmetric cosmology and then allowed the cosmologically relevant observational data to determine the solutions to the field equations. This has been done using the usual 3 + 1 LTB coordinates and showing in detail how the observations can be used to determine the space-time metric by numerical integration of the field equations. Others, including ourselves, have shown, alternatively, how to integrate the general spherically symmetric field equations in observational coordinates (OC), in which the observer's past light cone plays a prominent role. In fact, the radial coordinate in this framework is a null parameter measuring ``distance'' down the generators of the observer's past light cone.\\

In this paper, we shall discuss how the discrete observational data, with
its gaps and uncertainties, can be handled --  either by keeping it discrete and binning it by
redshift or by smoothing it into ``data functions'';  whether there are any difficulties with coordinate singularities at the maximum of the angular-diameter (observer area) distance, which is one of the key components of the observational data; and whether or not there are stability problems
with the equations themselves. Will small uncertainties in the data lead to a ``blow up''
of the solutions we are seeking? Careful analysis of these aspects of the OC integration
procedure will allow us to compare and contrast it with the 3 + 1 LTB approach \cite{mustaph:hell, liu:hell, mcclur:hell}, and to assess the advantages and disadvantages of each. At the same time, we shall compare how the data can be used -- particularly the maximum of the angular-diameter distance -- in the two approaches to determine the value of the cosmological constant $\Lambda$
 for the more general LTB -- not just for FLRW -- models.\\
 
In the next section we define observational coordinates, write the general spherically symmetric metric using them and present the central conditions for the metric variables. Section \ref{sec:obspar} presents the basic observational parameters we shall be using and
several key relationships among the metric variables. Section \ref{sec:fieqs} presents the full set of field equations for the spherically symmetric case, with dust and with $\Lambda \neq 0$. In Sec. \ref{sec:nonflat}, we present a general integration scheme for all inhomogeneous spherically symmetric LTB models. Section \ref{sec:compa} is dedicated to  the comparison between the the OC formulation and integration and that of the usual $(3+1)$ approach. In Sec. \ref{sec:stability} we examine the stability of the solutions to the null Raychaudhuri equation. Since that is the first step in solving the field equations with data on our past light cone, stability of its solutions is necessary for insuring the stability of the solutions to the other field equations. This complements the discussion of stability of the OC integration in the last paragraphs of 
Sec. \ref{sec:compa}. Sec. \ref{sec:idata} 
discusses the implementation of an algorithm for using the spherically symmetric OC field equations with data to determine the metric. Section \ref{sec:conclusion} is devoted to our conclusions.

The new results in this paper are found in Secs.\ref{sec:compa}, \ref{sec:stability}, and \ref{sec:idata}. They include an explicit and detailed comparison of the OC and the $3 + 1$ 
LTB approaches to integrating the spherically symmetric field equations, a 
discussion of advantages and disadvantages of solving the equations numerically in 
contrast to first fitting the data with smooth ``data functions'', a brief 
explanation of how the $3+1$ approach can use the maximum of the angular 
diameter distance to determine the cosmological constant $\Lambda$, a study of the 
stability of the OC formulation of the field equations, particularly the null 
Raychaudhuri equation, and a detailed example (in Sec. \ref{sec:idata}) of carrying out 
the full integration of the OC equations, using FLRW data functions. Here 
no stability problems are encountered. Sections \ref{sec:obsmet} -- \ref{sec:nonflat} 
provide a succinct, but detailed synthesis of the OC approach, the OC field equations, 
and their integration with data functions on the light cone, and then off it to obtain the time-dependence of the space-time metric functions.

\section{ \label{sec:obsmet}The Spherically Symmetric Metric in Observational Coordinates}

In observational coordinates $\{w,y,\theta,\phi\}$ the Spherically Symmetric metric takes
the general form:
\begin{equation}
ds^2=-A(w,y)^2dw^2+2A(w,y)B(w,y)dwdy+C(w,y)^2d\Omega ^2,  \label{oc}
\end{equation}
where $w$ is the time coordinate defined such that $w=$constant specifies 
a single past light cone along the observer's world line ${\cal C}$. Our past light cone, here and now is given by $w=w_0$. $y$ is the null radial coordinate measured down the null geodesics generating each past light cone centered on ${\cal C}$, which is given by $y=0$. $y$ increases 
as one moves down the past light cones from ${\cal C}$ and is chosen to be co-moving with the matter flow, that is $u^a{\partial_a}y=0$,  $u^a$ being the fluid 4-velocity satisfying $u^a=A^{-1}\delta _w^a$ and $u^au_a=-1$. $\theta$ and $\phi$ are simply the latitude and longitude of 
observation -- i.e. spherical coordinates based on a parallely propagated orthonormal tetrad 
along ${\cal C}$. 

Besides the freedom to choose $y$ co-moving, there are other coordinate freedoms  which preserve the form of the metric Eq. (\ref{oc}). In particular we can choose $w$ to be any time parameter we like along ${\cal C}$ -- that is, at $y=0$. This is usually done by choosing 
$A(w,0)$. Furthermore, we can choose $y$ to be any null distance parameter on an initial past light cone -- usually on $w=w_0$. Since $y$ is co-moving, that choice is then dragged off by the fluid flow onto other light cones. We shall use this freedom to set 

\begin{equation}
A(w_0, y) = B(w_0, y).  \label{ab}
\end{equation}

For further details about observational coordinates see Ellis {\it et al} \cite{Ellis et al} and Ara\'ujo and Stoeger \cite{AS2009b,AS2010a}.

It is important to specify the central conditions for the metric variables $%
A(w, y)$, $B(w, y)$ and $C(w, y)$ in Eq. (\ref{oc}) -- that is, their proper
behavior as they approach $y = 0$. These are:
\begin{eqnarray}
{\rm as}\;\;y\rightarrow 0:\;\;\; &&A(w,y)\rightarrow A(w,0)\neq 0, 
\nonumber \\
&&B(w,y)\rightarrow B(w,0)\neq 0,  \nonumber \\
&&C(w,y)\rightarrow B(w,0)y = 0,  \label{cent} \\
&&C_y(w,y)\rightarrow B(w,0).  \nonumber
\end{eqnarray}
These conditions insure that ${\cal C}$, our world line, is
regular -- so that all functions on it our bounded, and that the
spheres ($w$, $y =$ const) go smoothly to ${\cal C}$ as $y \rightarrow 0$.
They also insure that the null surfaces $w =$ const are past light cones
of observers on ${\cal C}$ (See Ref.  \cite{Ellis et al} , especially section 3.2, p.
326, and Appendix A for details).

There are several other relationships which are important for later use. In particular we have
 two fundamental four-vectors in the problem, the fluid four-velocity $u^a$ and the null vector 
 $k^a$, which points down the
generators of past light cones. These are given in terms of the metric
variables as
\begin{equation}
u^a = A^{-1}\delta^a{}_w ~,~~ k^a = (AB)^{-1}\delta^a{}_y.  \label{uk}
\end{equation}
From the normalization condition for the fluid four-velocity, we
can immediately see that it can be given (in covariant vector form) as the
gradient of the proper time $\tau$ along the matter world lines: $u_a=-\tau,_a$.
It is also given by (\ref{oc}) and (\ref{uk}) as
\begin{equation}
u_a=g_{ab}u^b=-Aw_{,a}+By_{,a}.
\end{equation}
Comparing these two forms implies
\begin{equation}
d\tau=Adw-Bdy~~\Leftrightarrow~~A=\tau_w~,~~ B=-\tau_y,  \label{tw}
\end{equation}
which shows that the surfaces of simultaneity for the observer are given in
observational coordinates by $A dw = B dy$. 
The integrability condition of Eq. (\ref{tw}) is simply then

\begin{equation}
A^{\prime}+\dot{B}=0.  \label{coneq}
\end{equation}
where a ``dot'' indicates $\partial/\partial w$ and a ``prime'' indicates $\partial/\partial y$. 
This turns out precisely to be the momentum conservation equation, which is
a key equation in the system and essential to finding a solution. 

\section{\label{sec:obspar}The Basic Observational Quantities}

The basic observable quantities on $ {\cal C}$, which are consistent with the central conditions,
are the following: \\

(i) Redshift. The observed redshift $z$ at time $w_0$ on $ {\cal C}$ for a comoving source a
null radial distance $y$ down $C^{-}(p_0)$ is given by
\begin{equation}
1+z={\frac{A(w_0,0)}{A(w_0,y)}}.  \label{z}
\end{equation}

 (ii) Redshift-drift (time-drift of the cosmological redshift).  In principle we can measure the 
 change in the cosmological redshift of distant galaxies from one time of observation to another 
(as a function of redshift). This is the redshift drift $\dot{z}(z)$. As a function of $y$, using
Eq. ({\ref{z}}), it can be written

\begin{equation}
\dot{z}(w_0, y) = (1+z)\Biggl\{\frac{\dot{A}(w_0,0)}{A(w_0,0)} -
\frac{\dot{A}(w_0,y)}{A(w_0,y)} \Biggr\}. 
\end{equation}\\

As many others have shown, and we have confirmed within the OC framework, [ \cite{sand, mcv, uce, AS2010a}; see also Appendix \ref{sec:ap1} ]  $\dot{z}$ contains evolution- and
galaxy-count-independent information about the mass-energy density. With reliable 
measurements of $\dot{z}$ and of observer-area
(angular-diameter) distances, we can determine the mass-energy density
$\mu(w_0,z)$. This would enable us to avoid having to use galaxy number counts
(see below) to determine that. It turns out, besides, that $\dot{z}$ as a
measurement on our past light cone can be used to determine the important
relationship $z = z(y)$, which enables us to translate key data as functions of
$z$ into functions of $y$ [see Appendix \ref{sec:ap1}]. Without $\dot{z}$ data, as we shall
see below, $z = z(y)$ can be determined only by solving the null Raychaudhuri
equation [Eq. (\ref{nr}) below], which needs, instead, reliable galaxy-number-count
and average galaxy mass data, which is very difficult to obtain.

Though we do not yet have the technological capability to measure $\dot{z}$,
the prospects for developing it over the next 20 years are good. Loeb \cite{Loeb} and
more recently Pasquini {\it et al.} \cite{Pasq} have studied in depth the requirements
and the promising ways of doing so, using the hydrogen and metal absorption
lines in Ly-$\alpha$ forests. Pasquini, {\it et al.} \cite{Pasq} using simulations of the spectra 
along with analysis of the expected noise, as well
as the projected precision and stability of the planned European Extremely Large
Telescope (E-ELT) and its CODEX ultra-stable spectrograph, conclude that a
measurement of redshift drift with significant precision would be possible with
these instruments over a ten-year time-line [ see also Dunsby {\it et al.} \cite{Dunsbyetal} ].\\

(iii) Observer area distance (Angular diameter distance). The observer area distance, often written as $r_0$, measured at time $w_0$ on $ {\cal C}$ for a source at a null radial distance $y$
is simply given by
\begin{equation}
r_0=C(w_0,y),
\end{equation}
provided the central condition (\ref{cent}), determining the relation
between $C(w,y)$ and $B(w,y)$ for small values of $y$, holds. This quantity
is also measurable as the luminosity distance $d_L$ because of the reciprocity
theorem of Etherington \cite{Etherington33} (see also Ellis \cite{Ellis 1971}),\\ 

\begin{equation}
d_L = (1+z)^2 C(w_0, y).   \label{recth} \\
\end{equation}

(iv) The maximum of observer area distance. Generally speaking, $C(w_0, y)$
reaches a maximum $C_{max}$  for a redshift $z_{max}$ between about $1.6$ and $2.0$ 
[Hellaby  \cite{Hellaby}; see also Ellis and Tivon  \cite{ET} and Ara\'{u}jo and Stoeger  \cite{ASII} ]. At
$C_{max}$, of course, we have 

\begin{equation}
\frac{d C(w_0, z)}{d z} = \frac{d C(w_0, y)}{d y} = 0,\\
\end{equation}
further conditioned by
\begin{equation}
\frac{d^2 C(w_0, z)}{d z^2} < 0.\\
\end{equation}
These $C_{max}$ and $z_{max}$
data provide additional independent information about the cosmology. Without
$C_{max}$ and $z_{max}$ we cannot constrain the value of $\Lambda$ with our other data.
 
(v) Galaxy number counts. The number of galaxies counted by a central
observer out to a null radial distance $y$ is given by
\begin{equation}
N(y)=4\pi\int_0^y \mu(w_0,\tilde{y})m^{-1}B(w_0,\tilde{y})C(w_0,\tilde{y})^2
d\tilde{y},  \label{N}
\end{equation}
where $\mu$ is the mass-energy density and $m$ is the average galaxy mass.
Then the total mass-energy density can be written as
\begin{equation}
\mu(w_0,y) = m\;n(w_0,y) = M_0(z)\;{\frac{dz}{dy}}\;{\frac{1}{B(w_0,y)}},
\label{mudef}
\end{equation}
where $n(w_0, y)$ is the number density of sources at $(w_0, y)$, and where
\begin{equation}
M_0 \equiv {\frac{m}{J}}\;{\frac{1}{d\Omega}}\;{\frac{1}{r_0^2}}\;{\frac{dN}{dz}}. \label{m0def}
\end{equation}
Here $d \Omega$ is the solid angle over which sources are counted, and
$J$ is the completeness of the galaxy count, that is, the fraction of
sources in the volume that are counted is $J$. The effects of dark
matter in biasing the galactic distribution may be incorporated via $m$ and/or
$J$ .  In order to effectively use number counts to constrain our cosmology, we shall also
need an adequate model of galaxy evolution. We shall not discuss this
important issue in this paper. But, fundamentally, it would give us
an expression for $m = m(z)$ in Eqs. (\ref{mudef}) and (\ref{m0def}) above.

In line with our discussion of redshift drift earlier in this section, Ara\'ujo and Stoeger \cite{AS2010a}  have recently shown within the OC framework that the mass-energy density of the Universe can be fully determined in terms of the cosmological redshift, the time drift of the cosmological redshift, and the observer area distance [see Appendix \ref{sec:ap1}]. Using the redshift-drift as a replacement for mass-energy density element in the minimally required data set needed to construct an LTB model for the Universe means that the number counts data is no longer needed. Moreover, as pointed out by Liske {\it et al.} \cite{Liske} and Linder \cite{Linder} one of the main features
of  the redshift drift measurement is its model independency.  

In identifying, choosing and discussing these basic observable cosmological quantities or probes in the general spherically symmetric OC context, we are not intending to exclude the potential important contribution of other probes that are being studied and show promise --including weak gravitational lensing, baryonic acoustic oscillations, various cosmic microwave background radiation (CBR) measurements, and CBR-related measurements of the integrated Sachs-Wolfe( ISW) and the Sunyaev-Zel'dovich (SZ) effects. These essentially provide observational avenues, like those we have discussed here, to determining the mass-energy density of the universe, angular-diameter distances, etc. We have not included discussion of them here, because they are not as straight-forwardly related to cosmological parameters in the general OC context, and often rely, at least in many applications so far, on assuming that the universe is FLRW -- or perturbed FLRW -- on the largest scales.

\section{\label{sec:fieqs}The Spherically Symmetric Field Equations in 
Observational Coordinates}

Using the fluid-ray tetrad formulation of the Einstein's equations 
developed by Maartens \cite{m} and Stoeger {\it et al} \cite{fluid ray}, 
one obtains the spherically symmetric field equations in observational 
coordinates with $\Lambda \neq 0$ (see Stoeger {\it et al} \cite{OC III} for a
detailed derivation). Besides the momentum conservation Eq. (\ref{coneq}), they are
as follows:

A set of two very simple fluid-ray tetrad time-derivative equations:

\begin{eqnarray}
\dot{\mu}_m &=& -2{\mu_m} \left(\frac{\dot B}{2B} + \frac{\dot{C}}{C} \right), \label{mueqr} \\
\dot{\omega} &=& -3 \frac{\dot C}{C} \biggl(\omega + \frac{\mu_{\Lambda}}{6} \biggr), \label{omegaeqr}
\end{eqnarray}
where $\mu_m$ again is the relativistic mass-energy density of the dust, 
including dark matter, and 

\begin{equation}
\omega(w,y)\equiv -{\frac{1}{2C^2}} + {\frac{
\dot{C}}{{AC}}}{\frac{C^{\prime}}{{BC}}} + {\frac{1}{2}} \biggl({\frac{
C^{\prime}} {BC}}\biggr)^2, \label{omegadef}
\end{equation}
is a quantity closely related to $\mu_{m_0}(y)\equiv \mu_m(w_0,y)$ [see Eq.\ref{omegdef} below].

Eqs. (\ref{mueqr}) and (\ref{omegaeqr}) can be quickly integrated to give:
\begin{widetext}
\begin{equation}
\mu_m(w,y)=\mu_{m_0}(y)\;{\frac {B(w_0,y)} {B(w,y)}}\; \frac {C^{2}(w_0,y)} {C^{2}(w,y)}; 
\end{equation}

\begin{equation}
\omega(w,y)=\biggl(\omega_0(y) + {\frac {\mu_{\Lambda}} {6} }\biggr) {\frac{C^3(w_0, y)}{C^3(w,y)}}
- {\frac {\mu_{\Lambda}} {6}} = {-{\frac{1}{{2C^2}}}+{\frac{
\dot{C}}{{AC}}}{\frac{C^{\prime}}{{BC}}}+{\frac{1}{2}}\biggl({\frac{
C^{\prime}}{{BC}}}\biggr)^2}, \label{omega} 
\end{equation}
\end{widetext}
where $\omega_0(y)\equiv \omega(w_0,y)$ and the last equality in (\ref{omega}) follows from the definition of $\omega$ given above. In deriving and solving these equations, and those below, we have used the
typical $\Lambda$ equation of state, $p_{\Lambda} = - \mu_{\Lambda},$
where $p_{\Lambda}$ and $\mu_{\Lambda} \equiv \frac{\Lambda}{8 \pi G}$ are the pressure and the 
energy density due to the cosmological constant. Both $\omega_0$ and
$\mu_0$ are specified by data on our past light cone, as we shall show.
$\mu_{\Lambda}$ will eventually be determined from the measurement of $C_{max}$
and $z_{max}$. 

The fluid-ray tetrad radial equations are:
\begin{widetext}
\begin{eqnarray}
&{\frac{C^{\prime\prime}}{C}} = {\frac{C^{\prime}}{C}}{\biggl({\frac{
A^{\prime}}{A}} +{\frac{B^{\prime}}{B}}\biggr)} - {\frac{1}{2}}B^2\mu_m;
\label{nr} \\
&\biggl[ \bigl(\omega_0(y) + {\frac {\mu_{\Lambda}} {6} }\bigr)C^3(w_0, y)\biggr]^{\prime} = -{\frac{1}{2}}\mu_{m_0}\;
{B(w_0,y)}\;{C^{2}(w_0,y)}\;{\biggl({\frac{\dot {C}}{A}} +
 {\frac{C^{\prime}}{B}}\biggr)};  \label{omegap} \\
&{\frac{{\dot {C}}^{\prime}}{C}} = {\frac{{\dot B}}{B}}{\frac{C^{\prime}}{C}
} - \biggl(\omega + {\frac {\mu_{\Lambda}} {2}}\biggr)\; A\;B.  \label{prdot}
\end{eqnarray}
\end{widetext}
The remaining ``independent'' time-derivative equations given by the
fluid-ray tetrad formulation are:

\begin{eqnarray}
&&{\frac{{\ddot{C}}}C}={\frac{{\dot{C}}}C}{\frac{{\dot{A}}}A}+ \biggl(\omega + 
{\frac {\mu_{\Lambda}} {2}}\biggr) \;A^2;
\label{Cdd} \\
&&{\frac{{\ddot{B}}}B}={\frac{{\dot{B}}}B}{\frac{{\dot{A}}}A}-2\omega \;A^2-{%
\frac 12}\mu_{m} \;A^2 .  \label{bdd}
\end{eqnarray}
From Eq. (\ref{omegap}) we see that there is a naturally defined
``potential'' (see Stoeger {\it et al} \cite{OC III}) depending only on the radial
null coordinate $y$ -- since the left-hand-side depends only on $y$, the
right-hand-side can only depend on $y$:

\begin{equation}
F(y)\equiv {\frac{\dot{C}}A}+{\frac{C^{\prime }}B},  \label{f1}
\end{equation} 
Thus, from Eq.  (\ref{omegap}) itself
\begin{widetext}
\begin{equation}
\omega_0(y)= - {\frac {\mu_{\Lambda}} {6} }- {\frac{1}{2 C^3(w_0, y)}}\;\int{\mu_{m_0}(y)\;
{B(w_0,y)}\;{C^{2}(w_0,y)}\;F(y)\;dy}.  \label{omegdef}
\end{equation}
\end{widetext}
All these spherically symmetric equations in observational coordinates have
been independently obtained by Hellaby and Alfedeel \cite{hell:alna} directly from the
metric, without relying on the fluid-ray tetrad equations.

We now introduce a simple but very important observational relationship 
originally obtained by Hellaby  \cite{Hellaby} in the 3+1 framework [for a detailed
derivation of this result in observational coordinates see Ara\'ujo and Stoeger \cite{AS2009b}] :

\begin{equation} 
6M_{max} + \mu_{\Lambda} C^3_{max} - 3 C_{max} = 0 .\label{hellabyeq}
\end{equation}
 where, $M_{max}$ is the maximum of the quantity $M(y)$ given by

\begin{equation}
M(y) = \frac {1}{8 \pi} \int_0^y mN^{\prime}(\tilde y) F(\tilde y) d\tilde y  =  \frac {1}{8 \pi} \int_0^y \bar M^{\prime}(\tilde y) F(\tilde y) d\tilde y . \label{MN} 
 \end{equation}
 where, $ \bar M(y) = m N(y)$ is  the total mass summed over the whole sky by a central observer out to a null radial distance $y$.

Eq. (\ref{hellabyeq}) has to be considered a fundamental relation in Observational Cosmology, since it enables, from $C(w_0, z_{max})$ and $z_{max}$ measurements,  the determination of the unknown constant $\mu_{\Lambda}$.

Finally,  from Eq. (\ref{N})  we have 

\begin{equation}
mN^{\prime}(y) = 4 \pi \mu_{m_0}(y) B(w_0,y) C^2(w_0,y). \label{mNp}
\end{equation}
Hence, Eqs. (\ref{MN}) and (\ref{mNp}) give

\begin{equation}
M(y) = \frac{1}{2}\int_0^y \mu_{m_0}(\tilde y) B(w_0,\tilde y) C^2(w_0,\tilde y) F(\tilde y)d\tilde y.\label{Mmu}
\end{equation}
This clearly shows the relationship between $M(y)$ and the mass-energy density.

\section{\label{sec:nonflat}The General Solution - Time evolution off our Light Cone}

In this section we outline the general integration procedure that is applicable 
to all inhomogeneous spherically symmetric universe models  - that is the only constraint.  
[See Ara\'ujo and Stoeger \cite{AS2009b} for a detailed description of this procedure.]
We do not know whether the Universe is homogeneous  or not. But the data give us redshifts
$z$, observer area distances (angular diameter distances) $r_0(z)$,
the angular-diameter distance maximum $C_{max}(w_0, z)$ at $z_{max}$, and galaxy
number counts, which together with an overall average mass per galaxy
can in principle give us $M_0(z)$. Eventually with the advent of very large
telescopes and very high precision spectrographs, as we have already indicated, we may also have $\dot{z}(z)$
data, giving us the time drift of cosmological redshifts \cite{Loeb,Pasq}. 
It is important to specify $C_{max}(w_0, z)$ at $z_{max}$, because, as we
have already emphasized, without them, we do not have enough information
to determine all the parameters of the space-time in the $\Lambda \neq 0$ case.
Although we can determine $C(w_0, z)$ with good precision
(by obtaining luminosity distances $d_L$ and employing the reciprocity theorem,
Eq.  (\ref{recth})) out to relatively high redshifts, at present we do not yet have
reliable data deep enough to determine $C_{max}$ and $z_{max}$. But this is now within the realm of possibility, with precise space-telescope distance measurements for supernovae Ia, and the potential for the linear-size of ultra-compact (miliarcsecond) radio sources as a standard rod, enabling reliable $C(w_0,z)$ out to redshifts $z\approx 4$ (such data already exists, but the status of these sources as standard measuring rods must be better substantiated) \cite{jackson}. \\

In pursuing the general integration with these data, we use the framework
and the intermediate results we have presented in Sec. \ref{sec:fieqs}. Obviously, one
of the key steps we must take now is the determination of the potential
$F(y)$, given by Eq. (\ref{f1}). 
This means we need to determine $C^{\prime}(w_0, y)$ and $\dot{C}(w_0, y)$, which we
now write as $C_0^{\prime}$ and $\dot{C}_0$, respectively. We also need
$A(w_0, y).$ We remember, too, that on $w = w_0$ we have chosen
$B(w_0, y) = A(w_0, y)$, which we have the freedom to do. \\

Clearly, $C_0^{\prime}$ can be determined from the $r_0(z) \equiv C(w_0,z)$
data, through fitting, along with the solution of the null Raychaudhuri equation, 
Eq. (\ref{nr}), which has the first integral

\begin{equation}
\frac{dy}{dz} = \frac{1}{A_0} (1+z)^2 \frac{dr_0}{dz} {\Biggl\{1 - \frac{1}{2}\int_0^z (1+\tilde z)r_0(\tilde z)M_0(\tilde z)d\tilde z\Biggr\}}^{-1}, \label{nrfint}
\end{equation}
to obtain $z =z(y)$ upon inversion (Stoeger {\it {et al}}  \cite{OC III}). 
As already mentioned, this enables us to write all of our data as functions of $y$ instead
of as functions of the redshift $z$. In Sec. \ref{sec:compa}, and particularly in Sec.\ref{sec:stability}, we shall discuss this important step in the integration procedure, focusing 
on the stability of its solutions. We determine $\dot{C}_0$ by solving Eq. (\ref{prdot}) for
$w = w_0$. Its solution, taking into account the appropriate boundary condition (central condition) is given by:

\begin{equation}
\dot C_0(y) = \frac {1}{2C_0(y)} \int_0^y \Biggl( A_0^2(\tilde y) - \frac {2 A_0^{\prime}(\tilde y) C_0^{\prime}(\tilde y)C_0(\tilde y)}{A_0(\tilde y)} -(C_0^{\prime}(\tilde y))^2-A_0^2(\tilde y)C_0^2(\tilde y)\mu_{\Lambda} \Biggr) d\tilde y. \label{dotC}
\end{equation}

This procedure enables us to determine $F(y)$, which obviously also depends 
on  $\mu_{\Lambda}$.  Our next step is to insert this result along with $N^{\prime}(y)$ -- the spatial variation of the galaxy number counts -- and the average mass per galaxy $m$ into 
Eq. (\ref{MN}) to determine the mass function $M(y)$. 
Next we evaluate the mass function $M(y)$ at $y_{max}$ and plug the result into Eq. (\ref{hellabyeq}) which becomes an algebraic equation for  $\mu_{\Lambda}$. With this determination of $\mu_{\Lambda}$, we know $\dot{C}_0(y)$ completely,
and can now determine $F(y)$ from Eq. (\ref{f1}). Furthermore, we observe from Eq. (\ref{omegadef})  that the quantity $\omega_{0} (y)$  is also completely determined at this stage. From here on, we can now follow the solution off 
$w = w_0$ for all $w$.

It is shown in Ara\'ujo and Stoeger \cite{AS2009b} that from Eqs. (\ref{coneq}) and (\ref{bdd}) 
one obtains the following equation for $\dot{A}(w_0, y)$:

\begin{equation}
\dot{A}_{0}(y) = A_{0}(y)\Biggl\{ \int_0^y\Biggl[2\omega_{0}(\tilde y) +{
\frac 12}\mu_{m_{0}}(\tilde y)\Biggl]{\;A_{0}}^2(\tilde y)\;d{\tilde y} + C_1\Biggr\}. \label{Adplc}
\end{equation}
where we have written $\dot{A}_0(y)$ and  $A_0(y)$  for  $\dot{A}(w_0,y)$
and $A(w_0,y)$ respectively. Since we can set $A(w,0)=1$ [see Appendix \ref{sec:ap2}] , it is obvious that  $C_1 = \dot{A}_{0}(0)/A_{0}(0)=0$.

Hence, we have shown that the data on our past light cone determines $\dot{A}_{0}(y)$. 
Since we know $\dot{A}_{0}(y)$ from the data,  Eqs.  (\ref{Cdd}) and  (\ref{bdd}) evaluated on our past light cone become algebraic equations for  $\ddot{C}_{0}(y)$  and $\ddot{B}_{0}(y)$, respectively

\begin{eqnarray}
\ddot{C}_{0}(y) &=&  \frac{\dot{C}_0(y){\dot{A}_0}(y)}{A_0(y)}+\Biggl[\omega_{0}(y) +{
\frac 12}\mu_{\Lambda}(y)\Biggl]{\;A_{0}}^2(y)C_0(y) \label{Cddpnc} \\
\ddot{B}_{0}(y) &=& - \frac{A_0^{\prime}(y){\dot{A}_0}(y)}{A_0(y)}-\Biggl[2\omega_{0}(y) +{
\frac 12}\mu_{m_{0}}(y)\Biggl]{\;A_{0}}^2(y)B_0(y)  \label{Bddpnc}
\end{eqnarray}
where in the later we have used Eq. (\ref{coneq}). 

Ara\'ujo and Stoeger \cite{AS2009b} have shown that the next step in this procedure is to 
obtain an equation for $\ddot A_0(w_0,y)$ through differentiation of  Eq. (\ref{bdd}) with 
respect to $w$,  and use of Eq. (\ref{coneq}). That leads to:

\begin{eqnarray}
\ddot A_0(y)&=&-A_0(y) \int^y_0 \Biggl\{\frac{\ddot B_0(\tilde y) \dot A_0(\tilde y)}{A^2_0(\tilde y)} -\frac{\dot B_0(\tilde y)  \dot A^2_0(\tilde y)}{A^3_0(\tilde y)}  \nonumber\\
&  & \hfill{\qquad}  -\frac{1}{A_0(\tilde y)} \biggl\{\frac{\partial}{\partial w}\biggl[ \biggl(2\omega +{\frac 12}\mu_{m} \biggl)\;BA^2\biggl]\biggl\}_0 d \tilde y \Biggr\}-A_0(y)C_2 , \label{Addplc}
\end{eqnarray}
where, for the same reason as above, $C_2= \ddot A_0(0)/A_0(0)=0$. 

It is important to note that all quantities on the right-hand side of the above equation are obtainable either directly from the data or from the algorithmic steps in the procedure we are
outlining here [see Ara\'ujo and Stoeger \cite{AS2009b} for a detailed description of the algorithm]. Therefore, we have shown that  we can obtain  $\ddot A_0(y)$ from the data. It is clear now that repetition of this procedure will give us all time derivatives
of A, B and C on our past light cone, which means that $A(w,y)$ $B(w,y)$ and $C(w,y)$ are completely determined by data on our past light cone, and calculable as Taylor series. 

We see from the above procedure that each step begins by finding the successive time derivatives of the metric function $A(w,y)$ on our past light cone, $\partial^n_wA(w_0,y)$,
which are completely determined given our choice of the time coordinate $w$ as measuring proper time along our world line ${\cal C}$.

\section{\label{sec:compa} Comparison of the Observational-Cosmology (OC) and Orthogonal $(3+1)$ (MH) Integrations}

The formulation of general spherically symmetric cosmologies in OC coordinates
we have just presented (and discussed more fully elsewhere \cite{AS2009b,AS2010a}) is equivalent to that which Bondi \cite{Bondi}, Bonnor \cite{Bonnor}, and more recently  Mustapha, Hellaby and Ellis \cite{mustaph:hell}, Lu and Hellaby \cite{liu:hell} and  McClure and Hellaby \cite{mcclur:hell} have developed in orthogonal $(3 + 1)$ coordinates, where the radial coordinate is spacelike, instead of null, as it is in our OC treatment. Integration of the
equations with data in this earlier alternative formulation (which we shall from now
on refer to as the Mustapha-Hellaby (MH) formulation) of LTB models has been
developed and discussed in detail by Mustapha, Hellaby and Ellis (hereafter MHE) \cite{mustaph:hell}, and in much more detail by Lu and Hellaby \cite{liu:hell} and McClure and
Hellaby \cite{mcclur:hell}. Here we briefly sketch key aspects of that integration
and compare it with the one we have just discussed, emphasizing the common features
and the differences. This comprehensive comparison of the two approaches is one of the primary aims of this paper, and is new to the literature.\\

The MH formulation appears to be quite different to the equivalent OC formulation.
This stems principally from the fact that the MH form of the metric is expressed
in terms, as we have already mentioned, of a spacelike radial coordinate, which is
orthogonal to the time coordinate, whereas in OC the radial coordinate is null,
and measures distance down successive past-light cones of the observer. Thus,
MH sports a foliation of spacelike surfaces normal to the temporal axis, while
OC is characterized by a foliation of past light cones, which are, of course,
null surfaces, along the observer's world line -- a past light cone for each instant
of time. As mentioned in the OC literature, the principal reason for
this choice is to reflect the fact that all the observational data we receive
is really arrayed on the observer's past light cone. Furthermore, all the cosmological fluid
particles -- astronomically speaking, the galaxies, like the observer's own galaxy -- 
move along world lines which intercept the past light cones at different locations
along it -- that is, at different redshifts and at different angular positions on the plane
of the sky. So, the OC formulation is more natural. More importantly, transforming from 
OC to MH coordinates, or vice-versa, presents difficulties. There is in general no exact solution of the null geodesic equation for relating a spacelike radial coordinate to 
a null radial coordinate. There are also a number of other very convenient features of
the OC formulation (see below), despite some conceptual challenges. \\

A less obvious difference between the two formulations and integration procedures is
that in the MH framework the emphasis is on solving for the three arbitrary functions in
the system of field equations and exploring how the evolution of the space-time is determined
by them (see \cite{hell:alna} and references therein). In the OC approach the emphasis is on using the observational data directly to determine the metric functions, whose precise relationship to the data is highlighted and worked out before carrying out the integration. Although MHE recommend fitting the relevant data with smooth data functions, \cite{mustaph:hell} in the most recent detailed implementations of the MH case \cite{liu:hell,mcclur:hell} the approach is discrete, reflecting the discreteness of the data and the need for numerical integration of the field equations. \\

In the OC approach, in contrast, though there is no explict avoidance of a numerical,
discrete approach, we and our collaborators have envisioned
having analytic functional forms which the discrete data obeys, with the possibility of 
carrying out an analytic integration of the field equations, at least in some simple cases. This is
driven by the realization that the observer-area-distance-redshift and galaxy-number-count (or, equivalently, the mass-energy-density-redshift) data for an exact FLRW universe, whether with or without a cosmological constant $\Lambda$, do take theoretically derivable forms with 
definite characteristics -- and can be analytically integrated, as we show here and elsewhere [see \cite{AS,AABFS,ASAB} and references therein]. In fact, as has been also shown,
a space-time is FLRW if and only if that data can be exactly fit by one or other of those functional 
forms [see \cite{AS,ASAB} and Sec. \ref{sec:idata} below]. In practice this approach, of course, necessitates
fitting suitable smooth analytic functions to discrete data, after correcting them, and keeping
track of the deviations of the actual corrected data from the data functions.
Another reason for doing this is to illustrate more clearly the actual integration algorithm -- the sequence in which the steps in the algorithm must be implemented -- and to provide a check on possible later numerical integration procedures. Then if some or all of the steps in a given case  need to be numerically integrated, that can be done against that informed context. \\ 

Even with these strong differences, however, there are clear similarities, as there must be.
Among these, and one of the most prominent, is the need to solve the null Raychaudhuri equation, 
Eq. (\ref{nr}), and more immediately its first integral Eq. (\ref{nrfint}) -- in order to
obtain $z = z(y)$ or $z = z(r)$ for the OC and the MH approaches, respectively [$r$
is the spacelike radial coordinate in the MH formulation]. This key relationship is needed, because
all the data is given as functions of the redshift $z$, and must be rewritten as functions of either
$y$ or $r$, respectively. The solution scheme for this equation is almost the same in both
formulations [compare for instance Eq. (34) of Stoeger, et al. \cite{OC III} with 
Eq. (51) of MHE \cite{mustaph:hell} -- notice also that in neither version is there a singularity in the solution
of this equation, at the maximum of the observer-area distance (see
below)]. However, later work by Hellaby and coworkers
somewhat obscures this similarity. Interestingly enough, too, in both cases (generalizing the MH
treatment to the $\Lambda \neq 0$ case)  this equation is independent of 
the cosmological constant $\Lambda$. This facilitates the integration. \\

With determination of $z= z(y)$, or in the MH case $z=z(r)$, the data functions for $C_0(z) = r_0(z)$,
$N(z)$ or $M_0(z)$, and in the future $\dot{z}(z)$, can, as already indicated, be written as functions of $y$ or of $r$ on our past light cone in each approach. This is another obvious feature common to the both formulations. However, in the OC case knowledge of $z = z(y)$ immediately enables us to write down the $g_{00}$ component of the metric, $A_0(y) = A(w_0,y)$, on our past light cone $w = w_0$ by Eq. (\ref{z}). The freedom we have in choosing $y$ also enables us to write $B(w_0,y)$ by Eq. (\ref{ab}), and the data then gives us the other metric variable on $w_0$ as well, $C_0(y) = C(w_0,y) = r_0(z(y))$. Thus we quickly have the solution of the field equations on our past light cone, $w = w_0$. The next and more difficult part of the integration is to move this solution off $w_0$ into the past. \\

Accomplishing this requires, as seen in the two previous sections of this paper, determining the time-derivatives of the metric variables on $w_0$, as well as $F(y)$ and $M(y)$. Once we have $z = z(y)$ from Eq. (\ref{nrfint}), and have written the data as functions of $y$, we solve Eq. (\ref{dotC}) for $\dot{C}_0(y)$. This enables us to calculate $F(y)$, and then $M(y)$, via Eqs. (\ref{f1}) and (\ref{MN}), respectively. We already know $\dot{B}_0(y)$ from Eq. (\ref{coneq}) $\Bigl[ \dot{B}_0(y) = -A^{\prime}_0(y) \Bigr]$. As we have also just seen, $\dot{A}_0(y)$ can be determined via Eq. (\ref{Adplc}), and higher time derivatives of $A$, $B$ and $C$ are found through Eqs. (\ref{Addplc}), (\ref{Bddpnc}), (\ref{Cddpnc}), and their higher-order generalizations. These enable us to construct a Taylor-series solution in the time $w$ for all the metric variables. In carrying this out, there are no inherent calculational difficulties or singularities. $C^{\prime}(y) = 0$ at $y = y_{max}$, but this quantity never appears in a denominator -- and so by itself never induces a singularity or an instability at $y_{max}$. Furthermore, the actual equations we are employing to find the solutions are at this stage linear. Once we know $M(y)$, of course, as we have already emphasized, we can, using our values of $C_0(y_{max})$ from the data, determine $\mu_{\Lambda}$ via Eq. (\ref{hellabyeq}). \\
    
The MH case is somewhat different, particularly with regard to induced singularities and instabilities due to $d \hat{R}/{dr} = C^{\prime}_0(y) = 0$ at $z = z_{max}$, which in the MH formulation now appears in the denominator of the some of the key equations -- though not in MHE's solution to the null Raychaudhuri equation (\ref{nr}), as we pointed out above. $\hat{R}(r)$ is the MHE designation for the angular-diameter (observer-area) distance down our past light cone, for which we have used $C_0(y)$ in our OC treatment. Once $z = z(r)$ is calculated and the data are expressed
as functions of $r$, $M(r)$ must be found by solving the differential equation (see MHE, Eq. (14)):

\begin{equation}
\frac{dM}{dr}+\Biggl(\frac{4\pi\hat{\rho}\hat{R}}{d\hat{R}/dr}\Biggr)M = \Biggl(\frac{2\pi\hat{\rho}\hat{R}^2}{d\hat{R}/dr}\Biggr)
\Biggl[\Biggl(\frac{d\hat{R}}{dr}\Biggr)^2 + 1 - \frac{1}{3}\Lambda \hat{R}\Biggr], \label{MHE14}
\end{equation}
where we have included the $\Lambda$ term which MHE set to zero in their treatment. Here we see clearly the radial derivative of $\hat{R}$ in the denominators, which induces a singularity in these terms at the maximum of $\hat{R}$. Once $M(r)$ is determined, $E(r)$, which in the MH formulation gives the local energy per unit mass of the cosmological fluid (which is essentially carries the same information as the potential $F(y)$ in the OC formulation), is found via the algebraic relationship

\begin{equation}
1 + 2E = \frac{\frac{1}{2}\biggl[\Bigl(\frac{d\hat{R}}{dr}\Bigr)^2 + 1\biggr] - \frac{M}{\hat{R}}}
{\Bigl(\frac{d\hat{R}}{dr}\Bigr)^2}. \label{Eequation}
\end{equation}
Obviously, once $M(r)$ is known we can solve Eq. (\ref{hellabyeq}) with $M = M_{max}$ and $C_{max} = \hat{R}_{max}$ as an algebraic equation for $\Lambda$. This complements the MH formulation already in the literature by showing how $\Lambda$ can be determined in that approach, using the maximum of the angular-diameter distance. $M(r)$ and $E(r)$ are two 
of the three arbitrary functions which must be determined in order to obtain the solution in the MH scheme -- the ``the bang time'' $t_B (r)$ is the other. In our OC framework, we, instead, focus on the equivalent problem of using the data to exactly determine the metric variables
as functions of $w$ and $y$.\\

The singularities in Eqs. (\ref{MHE14}) and (\ref{Eequation}), and in other equations of the MH formulation of LTB cosmologies, present a challenge to the integration scheme. In their numerical integration of the equations Hellaby and Lu \cite{liu:hell} and Hellaby and McClure \cite{mcclur:hell} overcome these difficulties by carefully matching the solutions on either side of this boundary. As mentioned above, this problem does not arise in the OC formulation. There, as we have seen, the equivalent $C_0^{\prime}(y) = 0$ at $y_{max}$ never appears in the denominator. This difference also significantly differentiates the two stability analyses as well. In the next section, we present a brief analysis of the stability of the null Raychaudhuri equation in its OC formulation. As we recall, solution of this equation is the first step in the OC integration process. 
This compares favorably with that in the MH formulation, where again singularities are involved \cite{mcclur:hell}.\\

Though there are some good reasons to follow the Hellaby and Lu, and the Hellaby and McClure, preference for avoiding fitting the data (e.g. angular-diameter distances and galaxy number counts as functions of redshift $z$) with smooth data functions, there are equally good reasons for doing so, as we have briefly indicated earlier. Of course, as in the MH approach, we need first to purify the actual data by removing errors due to selection effects, absorption, proper motions, image distortions, etc. Once that is done, we find the best functional fit to the corrected data using a function which has the proper overall characteristics that parameter (e.g. angular-diameter distance) must manifest, in terms of limiting values (e.g. at $z = 0$), maxima and minima, inflection points, etc. These are determined from the basic theory underlying the definition of the parameter, and the cosmological model itself, and severely limit the type of functions we can use. Furthermore, using smooth data functions fit to the observed data minimizes problems stemming from redshift-binning of data in the fully discrete approach. Finally, as we have pointed out above, having at least a formally analytic solution -- particularly in simple cases -- can illuminate and guide our numerical work in more complicated ones, and provide a way of validating it. Now we turn to examine the stability of the null Raychaudhuri equation.\\

\section{\label{sec:stability} Stability of the Null Raychaudhuri Equation}

As uncertainties and errors in the data are introduced into the OC field equations, how are the solutions affected? Will they experience any  ``blow-ups'' from small or modest errors or uncertainties?

We have already seen in the last section that we do not expect any -- at 
least not from singularities in the equations themselves. But we now, as an 
example, look more carefully at the first step of the integration procedure, 
the solution of the null Raychaudhuri equation [see Eq. (\ref{nr}) above], to see 
if it is stable against blow-ups induced by small errors or uncertainties. We 
recall that its solution enables us to find $z = z(y)$, which will enable us to 
write other cosmological functions in terms of $y$ instead of $z$. 

For our purposes in this section it is convenient to write the first integral of the 
Raychaudhuri equation, Eq. (\ref{nrfint}), as

\begin{equation}
\phi(z,C_0(z),M_0(z)) = \frac{(1+z)^2 C_{0,z}(z)}{A_0\Biggl[1 -\frac{1}{2}\int_0^z(1+\tilde z)C_0(\tilde z)M_0(\tilde z) d\tilde z\Biggr]}\label{nrfirst}
\end{equation}
where, $\phi \equiv dy/dz$. The change in $\phi$ due to uncertainties and errors in the data is
given by

\begin{equation}
\Delta \phi = \frac{\partial \phi}{\partial C_0} \delta C_0 +\frac{\partial \phi}{\partial M_0} \delta M_0 \label{dphi} 
\end{equation}
and, it follows from Eq. (\ref{nrfirst}) that

\begin{eqnarray}
\frac{\partial \phi}{\partial C_0}&=&A_0I(z)C_{0}(z)M_{0}(z)\label{phic}\\
\frac{\partial \phi}{\partial M_0}&=&A_0I(z)C_{0}(z)M_{0}(z)\frac{C_{0,z}(z)} 
{M_{0,z}(z)} \label{phim}
\end{eqnarray}
where,
\begin{equation}
I(z)\equiv \frac{ (1+z)^3}{2A_0^2\Biggl[1 -\frac{1}{2}\int_0^z(1+\tilde z)C_0(\tilde z)M_0(\tilde z) d\tilde z\Biggr]^2} \label{iz}
\end{equation}

It is clear from the above that the changes in $\phi$ will be small as long as $\delta C_0$ and
$\delta M_0$ are small. It follows from Eq. (\ref{phim}) that,  at $C_{max}=C(w_0,z_{max})$, 
${\partial \phi}/{\partial M_0}$ vanishes. Hence, 

\begin{equation}
(\Delta \phi)_{z_{max}} = A_0I_{max}C_{max}M_{max}\delta C_0 \label{dphicm} 
\end{equation}
where, $I_{max}\equiv I(z_{max})$ and $M_{max}\equiv M_0(z_{max})$.
Moreover, it is also clear here that there are no singularities that could cause divergences.

\section{\label{sec:idata} Determining a Universe Model Given Data Functions}

This section discusses the  implementation of the algorithm we presented
in \cite{AS2009b} for using the spherically symmetric OC field equations with data
to determine the metric.  [In Appendix \ref{sec:ap1} we briefly outline the OC integration 
when we are using redshift-drift data, rather than number counts, to gives mass-energy density.]
We assume that  the discrete observational data with
its gaps and uncertainties are smoothed it into  ``data functions''. For the sake of 
illustration, here we use a so-called ideal data set, that is data functions resulting
from a smoothing process modeling the Universe as being exactly flat FLRW
with  $\Lambda =0$.

The ideal FLRW data,  for the flat case with $\Lambda =0$, have the form 
(Stoeger, et al \cite{OC III})

\begin{equation}
C_0(z)=2H_0^{-1}(1+z)^{-2}\{z+1-(z+1)^{1/2}\} \label{cz}
\end{equation}
and 
\begin{equation}
M_0(z)=3H_0(1+z)^{1/2}, \label{mz}
\end{equation}
where $H_0$ is the Hubble parameter measured at $w=w_0$ and $y=0$.

Solving the null Raychaudhuri equation with this data (see Stoeger {\it et al%
} \cite{OC III}) yields the following relation between redshift and the null
coordinate $y$
\begin{eqnarray}
&&(1+z)={\frac 1{\left( 1+\alpha y\right) ^2}},  \label{zy} \\
&&\alpha \equiv {\frac{H_0A_0}2}=\frac {1}{w_0}.  \label{alpha}
\end{eqnarray}

From Eq. (\ref{ab}) and the substitution of Eq. (\ref{zy}) into Eqs.  (\ref{z}) and (\ref{cz}), 
we find that on our past light cone 

\begin{eqnarray}
A(w_0,y)&=& A_0(1- {\alpha}y)^2,  \nonumber \\
B(w_0,y)&=&A_0(1- {\alpha}y)^2,  \nonumber \\
C(w_0,y)&=&A_0y(1- {\alpha}y)^2. \label{funct}
\end{eqnarray}
respectively.

Application of the algorithm given in \cite{AS2009b} to data gives the time derivatives
of the metric functions $A(w,y)$, $B(w,y)$ and $C(w,y)$ of all orders evaluated on our 
past light cone $w=w_0$. For the ideal FLRW data given by Eqs. (\ref{cz}) and (\ref{mz})
the first few terms are

\begin{eqnarray}
\dot A_0(y)&=&\Biggl(\frac {2\alpha^2 y}{1-\alpha y} + C_1\Biggr)A_0(y), \nonumber  \\
\ddot A_0(y)&=&\Biggl[ \frac{\dot A_0^2(y)}{2} + \frac{4y\alpha^2(C_1-2\alpha)}{1-\alpha y} 
+\Biggl(C_2-\frac{C_1^2}{2}\Biggr)\Biggr] A_0(y),  \nonumber  \\
\dot B_0(y)&=&2A_0\alpha(1-\alpha y),  \nonumber \\
\ddot B_0(y)&=&2A_0\alpha[(C_1-\alpha) - \alpha y(C_1-2\alpha)],  \nonumber \\
\dot C_0(y)&=&2A_0y\alpha(1-\alpha y)  \nonumber \\
\ddot C_0(y)&=&2A_0y\alpha[(C_1-\alpha) - \alpha y(C_1-2\alpha)]  \label{deriv}
\end{eqnarray}
where, the constants $C_n$  are integration constants to be determined as explained  
in the sequel.

According to the procedure developed in \cite{AS2009b} we have that 
$A(w,y)$, $B(w,y)$ and $C(w,y)$ are written as a Taylor series in $w$. Hence,
in particular,

\begin{equation}
A(w,y) = A_0(y) + \dot A_0(y) (w-w_0) + \frac{1}{2} \ddot A_0(y)(w-w_0)^2 + ...
\end{equation}

On our world line $(y=0)$,  the above series reads

\begin{equation}
A(w,0) = A_0 [ 1+ C_1 (w-w_0) + \frac{1}{2} C_2 (w-w_0)^2 + ... ] \label{aw0}
\end{equation}

As explained in Sec. \ref{sec:obsmet} above,  the freedom we have to choose 
$w$ as any time parameter we like along our world line $\cal C$, is effected by choosing
$A(w,0)$. Hence, we clearly see from Eq. (\ref{aw0}) that the constants $C_n$ 
appearing in the integration procedure are not  completely arbitrary or independent. 
We are free to choose these constants as long as they satisfy the constraint  of being the coefficients in a Taylor expansion of $A(w,0)$. Next we illustrate this important conceptual
issue of the observational coordinates formalism by presenting two possible choices
of the time coordinate $w$ on our world line and their effects on the constants $C_n$.  

(i) Let us choose $w$ to measure proper time along our world line $\cal C$. That choice
[see  Appendix \ref{sec:ap2}  for details]  leads to 

\begin{equation}
A(w,0)=1, \label{awo2}
\end{equation}
and from  Eq. (\ref{aw0}) we find that $A_0=1$,  and all the constants $C_n$ vanish.

(ii) Consider the usual way in which the flat FLRW $(\Lambda = 0)$ metric  is written in
observational coordinates \cite{OC III},  \cite{AS}:

\begin{equation}
ds^2 = \frac{4}{(H_0w_0)^2}\Biggl(\frac{y-w}{w_0}\Biggr)^4\Big(-dw^2+2dydw+y^2d{\Omega}^2\Bigl) \label{flatFLRW}
\end{equation}
It follows from Eqs. (\ref{flatFLRW}) and (\ref{alpha}) that 

\begin{equation}
A(w,0)=A_0(\alpha w)^2 \label{aw03}
\end{equation}
Hence,  Eq. (\ref{aw0}) gives $C_1=2 \alpha$, $C_2=2\alpha^2$, $C_n=0$
for $n \geq 3$ and $A_0 \neq 0$. We note that the series is truncated at the third term.
Therefore, from the expressions (\ref{funct}) and (\ref{deriv}) we can reconstruct the solution
corresponding to the specific choice of time coordinate on our world line given by Eq. (\ref{aw03}).

\section{\label{sec:conclusion} Conclusion}

Here we have focused on key theoretical and mathematical aspects of
determining the large-scale metric of the universe from cosmologically 
relevant astronomical observations, comparing and contrasting the
traditional MH 3+1 LTB and the observational cosmology (OC) approaches
to formulating and integrating the inhomogeneous spherically symmetric
field equations with a cosmological constant, $\Lambda \neq 0$. Such
approaches are crucial, since,
despite the attractiveness and success of almost-FLRW models, there
are significant uncertainties about whether or not they provide the unique
best-fit description of the large-scale structure of the cosmos. These
uncertainties can
really only be resolved by non-perturbative treatments which go beyond
FLRW. In pursuing such paths we may find that we can confirm that, yes, 
the universe is indeed almost-FLRW
on the largest scales, and that the Copernican principle holds. In this case,
we would also be able to determine more precisely the smallest length-scale
on which the universe is almost-FLRW (it is clear from its small and intermediate
scale lumpiness that it is not almost-FLRW on those scales). However, a
second possible outcome of such studies is that the universe is not
almost-FLRW on the largest scales, or was not or will not always be so. These
approaches, beginning with spherically symmetric inhomogeneous models, would then
provide a more secure best-fit model to the observations, and a measure of 
how much deviation there is from the cosmological principle. More importantly
they would, as we have discussed in this paper, confirm whether or not there is
indeed a significant amount of vacuum energy, or dark energy, in the universe -- or
whether instead the apparent acceleration of cosmic expansion is due to 
large-scale inhomogeneities. 
We may end up confirming that the universe is almost-FLRW with 
$73\%$ dark energy, as the concordance cosmic model says, or we may find that
it is really close to an inhomogeneous LTB with or without dark energy. Either way, these
programs would significantly contribute to our cosmological knowledge and understanding, and
our level of confidence in it. \\

In the course of the paper we have summarized the essentials of both the OC and MH 
approaches -- the first employing observational coordinates suited to the past-light-cone
structure of our observations made here and now on our world line, and the second
relying on a 3 + 1 foliation of space-like surfaces orthogonal to the cosmic
time coordinate. In particular we sketched the essential steps in integrating the field equations
with data in each of the two formulations, high-lighting the main similarities and the differences.
In the OC case we also show -- in Appendix \ref{sec:ap1} -- how redshift drift can be used to
give us the mass-energy density of the universe as a function of redshift -- once we are able to 
obtain such data -- instead of the more problematic galaxy counts (this obviously can also
be done in the MH formulation). Although both formulations and integration procedures
promise to lead to the same results, the OC procedure manifests certain advantages --
particularly in the avoidance of coordinate singularities at the maximum of the
angular-diameter distance, and in the stability of the solutions obtained.
The OC procedure is, as already emphasized, geared towards determining the metric
variables from observations, whereas MH procedure is directed towards determining
the three free functions that appear in that formulation. These outcomes are equivalent, of course,
and each perspective has advantages. But the OC approach enables one to link the
metric more directly to observations. Finally, the possibility of performing some
analytic integrations, at least in simple cases, by first fitting the corrected data
to smooth data functions seems to have definite advantages to simply numerically
integrating the difference equations directly with binned data. Smoothed data
functions enable us to include properties that data must have within the 
model. \\

This, together with the OC procedure itself, is illustrated in Sec. \ref{sec:idata} with the 
very simple abbreviated example of deriving the FLRW metric from FLRW data. We have treated this before \cite{AS} by obtaining the solution on our past light cone. Here we 
completed that solution by moving it off the light cone, as specified in Sec. \ref{sec:nonflat}. We should note that, although FLRW data functions can be derived from the FLRW
metric, it is non-trivial -- and must be demonstrated -- that FLRW data functions
uniquely imply the FLRW metric. We suppose that we have ideal FLRW data, including knowing
that our universe is flat and that $z_{max} = 1.25$. We then carry out our integration
scheme for that data -- showing that it leads to a flat FLRW metric with $\Lambda = 0$.
This demonstrates the advantage of having data functions which can be used to construct
analytic solutions to the field equations. No singularity problems were encountered in 
carrying out this integration. We could have included uncertainties or errors in the
data functions, of course, and then determine how they propagate into the solution,
and whether or not they are large enough to affect its stability. This will be pursued later. \\

In a future paper we shall be investigate how cosmic microwave background (CMB) and
CMB anisotropy observations fit into this OC integration scheme, and to what extent
they give us information independent from other sources.

\appendix
\section{\label{sec:ap1}}

First, we briefly derive a relationship which enables us to find $z = z(y)$
using redshift-drift data -- without having to solve the null Raychaudhuri
equation, Eq. (\ref{nr}). 

It is clear that on the radial null geodesics $\tau = \tau (w,y)$ and $z=z(w,y)$. Now, since on our 
past light cone  $w=w_0$ -- in OC coordinates $w$ labels past light  cones -- we must have that

\begin{equation}
\frac{d\tau}{dz}=\frac{Adw - Bdy}{{\dot z}dw + \frac{\partial z}{\partial y}dy}, \label{dtdz2}
\end{equation}
where, $\dot z \equiv {\partial z}/{\partial w}$ is the redshift-drift. It follows that dividing the numerators and denominators on the right-hand side of  Eq. ({\ref{dtdz2}) by $dw$ gives $d\tau /dz=A/\dot z$. If, instead, we divide the numerators and denominators on the right-hand side  by $dy$ we obtain $d\tau/dz=-B({\partial y}/{\partial z})$. Equating these two results gives

\begin{equation}
\frac{B(w_0,y)}{A(w_0,y)}= -\frac{1}{\dot z}\frac{\partial z}{\partial y}
\label{jacob0}
\end{equation}
[See \cite{AS2010a} for a detailed derivation on this result.] Now, using the freedom to choose the radial null coordinate $y$ by setting $B(w_0,y) =A(w_0,y)$ [see Sec. \ref{sec:obsmet}],  Eq. (\ref{jacob0}) is further simplified on our past light cone.

Since ${\dot z(z)}$ is given from data, solving Eq. (\ref{jacob0}) gives $z=z(w_0,y)$. That is the same information that would be obtained from the null Raychaudhuri equation (\ref{nr}). 
It is quite clear that, whereas to solve the latter on our past light cone one needs the 
mass-energy density $\mu(w_0,z)$, what we need to solve Eq. (\ref{jacob0}) is a different piece
of information -- the redshift-drift  $\dot z(w_0,z)$.\\

Secondly, Ara\'ujo and Stoeger \cite{AS2010a} have shown that  using  Eq. (\ref{jacob0}) and its solution, we can write  the mass-energy density in terms of the redshift $z$ as

\begin{equation}
\mu_{m_0}(z) =   \frac {2\dot z_0(z)} {A^2_0(z)} \frac {\partial}{\partial z} \Biggl[ \frac {\dot z_0(z)}{1+z_0(z)}
\Biggl] -4\omega_0(z). \label{main6}
\end{equation}

Eq.  (\ref{main6}) shows that the mass-energy density $\mu_{m_0}(z)$ can be completely determined in terms of the redshift $z$, the redshift-drift $\dot z(z)$ and observer area 
distance $C(w_0,z)$ data on our past light cone. We observe that the $\mu_{m_0}(z)$
dependency on the observer area distance $C(w_0,z)$, leads to its dependency on 
$\mu_{\Lambda}$, that must be determined by data giving the maximum of  the observer area distance, $C_0(w_0, z_{\max})$, and the redshift $z_{max}$ at which that occurs.

Now, from Eqs.  (\ref{ab}),  (\ref{jacob0}) and (\ref{omegadef}) we find that

\begin{equation}
\omega_0(z) = -\frac{1}{2C_0^2(z)} \Biggl\{1+
\frac{ \dot z} {A^2_0(z)} \frac{dC_0(z)}{dz} \Biggl[2 \dot C_0(z) -  \dot z \frac{dC_0(z)}{dz}\Biggr]\Biggr\}
 \label{omega0}
\end{equation}
Clearly, $dC_0(z)/dz $ can be determined from the $r_0(z) \equiv C(w_0,z)$
data, through fitting. We determine $\dot{C}_0(z)$ by solving Eq. (\ref{prdot}) for  $\dot{C}_0(y)$ on $w = w_0$ [see Ara\'ujo and Stoeger \cite{AS2009b} for details] and then use Eq. (\ref{jacob0}) and its solution to write the result in terms of $z$. Taking into account the appropriate boundary condition (central condition),  $\dot{C}_0(z)$ is given by:

\begin{eqnarray}
\dot C_0(z) = - \frac {1}{2C_0(z)} & & \int_0^z \frac{1}{\dot z_0(\tilde z)}\Biggl \{A_0^2(\tilde z) 
- \dot z_0^2(\tilde z)\frac{dC_0(\tilde z)}{d \tilde z}
\Biggl[ \frac {2 C_0(\tilde z)}{ A_0(\tilde z)}\frac{dA_0(\tilde z)}{d \tilde z}+\frac{dC_0(\tilde z)}{d\tilde z}\Biggr] \nonumber \\
&  & \hfill{\quad}
 -A_0^2(\tilde z)C_0^2(\tilde z)\mu_{\Lambda} \Biggr\} d\tilde z. \label{coz}
\end{eqnarray}

Since, $\dot z(w_0,z)$ is very small, it follows very clearly from Eqs. (\ref{main6}), (\ref{omega0}) 
and (\ref{coz}) that a useful approximation to $\mu_{m_0}(z)$ is

\begin{equation}
\mu_{m_0}(z) \cong -4\omega_{0}(z).
 \label{main7}
\end{equation}

First, we note that having determined $\mu_{m_0}(y)$ from data, we can easily
find $M(y)$ by Eq. (\ref{Mmu}), once the potential $F(y)$ is known. But,
$F(y)$ is completely determined by data through Eqs. (\ref{f1}) and  (\ref{coz}), except
for the unknown value of the constant $\mu_{\Lambda}$, which we carry along for
the time being. Then we evaluate the mass function $M(y)$ at $y_{max}$ and plug
the result into Eq. (\ref{hellabyeq}), which with the $C_{max}$ data becomes an algebraic
equation for $\mu_{\Lambda}$. With this determination of $\mu_{\Lambda}$, we
know $\dot{C}_0(y)$ completely, and can now determine $F(y)$ more precisely
from Eq. (\ref{f1}) . Furthermore, we observe from Eq. (\ref{omegadef})  that the quantity
$\omega_0(y)$ is also completely determined at this stage. We are now ready to following 
our solution off out past light cone $w = w_0$ for all $w$ applying the procedure described in Sec. \ref{sec:nonflat} from Eq.  (\ref{Adplc}) onwards.

\section{\label{sec:ap2}}

Ellis {\it et al.} \cite{Ellis et al} have shown that if the coordinate $w$ is normalized by the condition that it measures proper time along $\cal C$, and $\tau$ denotes proper time along general galactic world lines, then the redshift observed at $\cal C$ for the light coming from a galactic world line is given by

\begin{equation}
1+z = \frac{dw}{d\tau} \label{ap1}
\end{equation}
with the ratio ${dw}/{d\tau}$ evaluated along the galactic world line.

Since $y$ is a comoving coordinate, and $d\tau^2\equiv -ds^2$, then Eqs.  (\ref{oc})
and (\ref{ap1}) lead to

\begin{equation}
1+z = \frac{dw}{d\tau}=\frac{1}{A(w,y)} \label{ap2}
\end{equation}
where the normalization at $y=0$ (our world line) gives

\begin{equation}
\Biggl(\frac{dw}{d\tau}\Biggr)_{y=0}=\frac{1}{A(w,0)}=1 \label{ap3}
\end{equation}


\begin{thebibliography}{99}

\bibitem{CM}  C. Clarkson and R. Maartens, {\it Inhomogeneity and the foundations of concordance cosmology},  
\textit{Class. Quantum Grav. {\bf 27}} (2010) 124008 [arXiv:1005.2165].

\bibitem{CS} R.R. Caldwell and A. Stebbins, {\it A Test of the Copernican Principle},
\textit{ Phys. Rev. Lett. {\bf 100}}  (2008) 191302 [arXiv:0711.3459].

\bibitem{CFL} T. Clifton, P. G. Ferreira and K. Land,
{\it Living in a Void: Testing the Copernican Principle with Distant Supernovae},
\textit{ Phys. Rev. Lett. {\bf 101}} (2008) 131302 [arXiv:0807.1443]. 

\bibitem{ZS} P. Zhang and A. Stebbins, {\it Confirmation of the Copernican principle at Gpc radial scale and above from the kinetic Sunyaev Zel'dovich effect power spectrum}, arXiv:1009.3967.

\bibitem{Ellis et al} G.F.R. Ellis, S.D. Nel, R. Maartens, W.R. Stoeger,
and A.P. Whitman, {\it Ideal observational cosmology},
 \textit{Phys. Rep. {\bf 124}} (1985) 315.

\bibitem{OC III}  W.R. Stoeger, G.F.R. Ellis, and S.D. Nel, {\it Observational cosmology. III. Exact spherically symmetric dust solutions},
 \textit{Class. Quantum Grav. {\bf 9}}, (1992) 509.

\bibitem{AS} M.E. Ara\'ujo and W.R. Stoeger, {\it 	
Exact spherically symmetric dust solution of the field equations in observational coordinates with cosmological data functions}, 
\textit{Phys. Rev. D {\bf 60}}, (1999) 104020.

\bibitem{ASR} M.E. Ara\'ujo, S.R.M.M. Roveda, and W.R. Stoeger, {\it Perturbed spherically symmetric dust solution of the field equations in observational coordinates with cosmological data functions},
\textit{Astrophys. J. {\bf 560}}, (2001) 7 [arXiv:gr-qc/0105001].

\bibitem{AABFS} M.E. Ara\'ujo, R.C. Arcuri, M.L. Bedran, L.R. de Freitas  and W.R. Stoeger, {\it Integrating Einstein Field Equations in Observational Coordinates with Cosmological Data Functions: Nonflat Friedmann-Lema"tre-Robertson-Walker Cases},
\textit{Astrophys. J. {\bf 549}}, (2001) 716.

\bibitem{ASAB} M.E. Ara\'ujo, W.R. Stoeger, R.C. Arcuri and M.L. Bedran, {\it Solving Einstein Field Equations in Observational Coordinates with Cosmological Data Functions: Spherically Symmetric Universes with Cosmological Constant},
\textit{Phys. Rev. D {\bf 78}}, (2008) 063513 [arXiv:0807.4193].

\bibitem{AS2009b} M.E. Ara\'ujo and W.R. Stoeger, {\it 	
Obtaining the time evolution for spherically symmetric Lemaitre-Tolman-Bondi models given data on our past light cone}, 
\textit{Phys. Rev. D {\bf 80}}, (2009) 123517; D  {\bf 81}, (2010) 049903(E) [arXiv:0904.0730].

\bibitem{AS2010a} M.E. Ara\'ujo and W.R. Stoeger, {\it Using Time Drift of Cosmological Redshifts 
to find the Mass-Energy Density of the Universe}
 \textit{Phys. Rev. D {\bf 82}}, (2010) 123513 [arXiv:1009.2783].
 
\bibitem{mustaph:hell} N. Mustapha, C. Hellaby and G. F. R. Ellis, {\it Large scale inhomogeneity versus source evolution: Can we distinguish them observationally?},
\textit{Mon. Not. Roy. Astron. Soc. {\bf 292}}, (1997) 817 [arXiv:gr-qc/9808079]. 

\bibitem{liu:hell} T. H.C. Lu and C. Hellaby, {\it 	
Obtaining the spacetime metric from cosmological observations},
\textit{Class. Quantum Grav. {\bf 24}}, (2007) 4107 [arXiv:0705.1060].

\bibitem{mcclur:hell} M. L. McClure and C. Hellaby, {\it 	
The Metric of the Cosmos II: Accuracy, Stability, and Consistency},
\textit{Phys. Rev. D {\bf 78}}, (2008) 044005  [arXiv:0709.0875].

\bibitem{hell:alna} C. Hellaby and A. H. A. Alfedeel, {\it Solving the Observer Metric},
\textit{Phys. Rev. D {\bf 79}}, (2009) 043501 [arXiv:0811.1676].

\bibitem{MaaMat} R. Maartens and D. R. Matravers, {\it Isotropic and semi-isotropic observations in cosmology},
\textit{Class. Quantum Grav. {\bf 11}},  (1994) 2693.

\bibitem{InhomUniv} R. Maartens, N. P. Humphreys, D. R. Matravers,  and
W. R. Stoeger, {\it Inhomogeneous universes in observational coordinates},
\textit{Class. Quantum Grav. {\bf 13}}, (1996) 253;  {\bf 13}, (1996) 1689(E)  [arXiv:gr-qc/9511045].

\bibitem{Hellamulti} C. Hellaby, {\it Multicolor observations, inhomogeneity and evolution},
\textit{Astron. Astrophys. {\bf 372}}, 357 (2001) 357 [arXiv:astro-ph/0010641].


\bibitem{sand}  A. Sandage, {\it The change of redshift and apparent luminosity of galaxies due to the deceleration of selected expanding universes},
 \textit{Astrophys. J. {\bf 136}}, (1962) 319.
 
\bibitem{mcv}  G. McVittie, {\it Appendix : The change of redshift and apparent luminosity of galaxies due to the deceleration of selected expanding universes},
 \textit{Astrophys. J. {\bf 136}}, (1962) 334.
 
\bibitem{uce} J.P. Uzan, C. Clarkson and G.F.R.Ellis, {\it 	
Time drift of cosmological redshifts as a test of the Copernican principle},
\textit{Phys. Rev. Lett. {\bf 100}}, (2008) 191303  [arXiv:0801.0068].

 \bibitem{Loeb} A. Loeb, {\it Direct Measurement of Cosmological Parameters from the Cosmic Deceleration of Extragalactic Objects},
\textit{Astrophys. J. {\bf 499}}, (19980 L111.

\bibitem{Pasq} L. Pasquini {\it et al.}, {\it CODEX: Measuring the Expansion of the Universe (and beyond)},
\textit{The Messenger {\bf 122}}, (2005) 10.

\bibitem{Dunsbyetal} P. Dunsby, N. Goheer, B. Osano and J. P. Uzan,  {\it 	
How close can an Inhomogeneous Universe mimic the Concordance Model?},
\textit{ JCAP {\bf 06}} (2010) 017 [arXiv:1002.2397].

 \bibitem{Etherington33} I.M.H. Etherington, {\it On the definition of distance in general relativity},
 \textit{Philos. Mag. ser. 7 {\bf 15}},  (1933) 761.

\bibitem{Ellis 1971}  G.F.R. Ellis, {\it Relativistic Cosmology},  in 
\textit{General Relativity and Gravitation}, R.K. Sachs ed.,  Academic Press, New York, (1971) p. 104.

\bibitem{Hellaby} C.W. Hellaby, {\it The Mass of the Cosmos},
\textit{Mon. Not. Roy. Astron. Soc. { \bf 370}}, (2006) 239 [arXiv:astro-ph/0603637].

\bibitem{ET}  G.F.R. Ellis and G. Tivon, {\it Observational relationships in inflationary universes and other cosmologies}, 
\textit{Observatory  { \bf 105}}, (1985) 189.

\bibitem{ASII} M.E. Ara\'ujo and W.R. Stoeger, {\it The angular-diameter distance maximum and its redshift as constraints on $\Lambda \ne 0$  Friedmann-Lema\^{\i}tre-Robertson-Walker models},
\textit{Mon. Not. Roy. Astron. Soc. { \bf 394}}, (2009) 438 [arXiv:0705.1846].


\bibitem{Liske} J. Liske {\it et al.}, {\it Cosmic dynamics in the era of extremely large telescopes},
\textit{Mon. Not. Roy. Astron. Soc. { \bf 386}}, (2008) 1192 [arXiv:0802.1532].
 
 \bibitem{Linder} E.V. Linder, {\it Mapping the Cosmological Expansion},
\textit{Rep. Prog. Phys. { \bf 71}}, (2008) 056901 [arXiv:0801.2968].

\bibitem{m} R. Maartens, {\it Idealized observations in relativistic cosmology},
Ph.D. thesis, University of Cape Town, (1980).

\bibitem{fluid ray}  W. R. Stoeger, S.D. Nel,  R. Maartens, and  G.F.R. Ellis, {\it The fluid-ray tetrad formulation of Einstein's field equations}, 
\textit{Class. Quantum Grav. {\bf 9}}, (1992) 493.

\bibitem{jackson} J. C. Jackson,  {\it 	Is there a standard measuring rod in the Universe?},
 \textit{Mon. Not. Roy. Astron. Soc. { \bf 390}}, (2008) L1 [arXiv:0810.3930].
 
\bibitem{Bondi} H. Bondi, {\it Spherically symmetric models in general Relativity}
 \textit{Mon. Not. Roy. Astron. Soc. { \bf 107}}, (1947) 410.

\bibitem{Bonnor}  W. B. Bonnor, {\it Evolution of inhomogeneous cosmological models}
\textit{Mon. Not. Roy. Astron. Soc. { \bf 167}}, (1974) 55.
 
\end{thebibliography}
\end{document}